\documentclass{IAU}
\usepackage{graphicx}

\title[Migration of bodies to the Earth]
{Migration of bodies to the Earth from different distances from the Sun}

\author[Sergei I. Ipatov]   {Sergei I. Ipatov}

\affiliation{Vernadsky Institute of Geochemistry and Analytical Chemistry of RAS, 
\\ 119991, 19 Kosygin st.,
Moscow, Russia \\ email: {\tt siipatov@hotmail.com}}

\pubyear{2022}
\volume{374}  
\jname{ Astronomical hazards for life on Earth}
\editors{Gonzalo Tancredi, ed.}

\begin{document}

\maketitle

\begin{abstract}
Migration of bodies under the gravitational influence of almost formed planets was studied, and probabilities of their collisions with the Earth and other terrestrial planets were calculated. Based on the probabilities, several conclusions on the accumulation of the terrestrial planets have been made.
The outer layers of the Earth and Venus could accumulate similar planetesimals from different regions of the feeding zone of the terrestrial planets. 
The probabilities of collisions of bodies during their dynamical lifetimes with the Earth could be up to 0.001-0.01 for some initial semi-major axes between 3.2 and 3.6 AU, whereas such probabilities did not exceed $10^{-5}$ at initial semi-major axes between 12 and 40 AU. 
The total mass of water delivered to the Earth from beyond Jupiter's orbit could exceed the mass of the Earth's oceans.
The zone of the outer asteroid belt could be one of the sources of the late-heavy bombardment.
The bodies that came from the  zone of Jupiter and Saturn 
typically collided with the Earth and the Moon with velocities from 23 to 26 km/s and  from 20 to 23 km/s, respectively.

\end{abstract}

\firstsection 

\section{Introduction}
Migration processes played a great role in accumulation of planets and in delivery of water to the Earth and the terrestrial planets. The life on the Earth could not appear without water. However, collisions of bodies with the Earth can be dangerous for life and can kill some organisms, including people. The Earth’s ocean water and its D/H ratio could be the result of mixing water from several exogenous and endogenous sources with high and low D/H ratio. 
It was considered that bodies containing ices migrated to the Earth from the outer part of the Main
asteroid belt (\cite[O'Brien et al., 2014;]{O'Brien2014} \cite[Morbidelli et al., 2000;] {Morbidelli2000} \cite[Morbidelli et al., 2012;] 
{Morbidelli2012} \cite[Petit et al., 2001;]{Petit2001} \cite[Raymond et al., 2004;] {Raymond2004}
\cite[Lunine et al., 2003]{Lunine2003}) 
and from beyond Jupiter's orbit 
(\cite[Morbidelli et al., 2000;]{Morbidelli2000}  \cite[Levison et al., 2001;]{Levison2001} \cite[Ipatov \& Mather, 2004;] {IpatovMather04}
\cite[Ipatov \& Mather, 2006;] {IpatovMather06}
\cite[Ipatov, 2010;] {Ipatov2010} \cite[Marov \& Ipatov, 2018)]{Marov2018}. 

The studies of the process of accumulation of the terrestrial planets were often (e.g,. \cite[Chambers \& Wetherill, 1998] {ChambersWetherill1998}; \cite[Ipatov, 1993]{Ipatov93}; \cite[Chambers, 2013]{Chamber2013}; \cite[Raymond et al., 2009]{Raymond2009}) based on computer simulations of the evolution of disks of gravitating bodies combined at their collisions. In (\cite[Ipatov, 1993]{Ipatov93}) initial planetesimals were divided into four groups depending on their semi-major axes. It was obtained that composition of the largest formed planets was similar and was close to the composition of the initial disk.

\section{Initial data and considered models}

Below migration of bodies-planetesimasl is considered at the stage when planets already have got their masses and orbits.
Such migration corresponds to the late stages of accumulation of planets and to the migration of bodies in the present Solar System. The main attention is paid to the probabilities of collisions of bodies with the Earth and the Moon. The studies were based on the computer N-body simulations of migration of bodies-planetesimals under the gravitational influence of planets. For integration of the motion equations, the symplectic method of the Swift integration package \cite[(Levison \& Duncan, 1994)]{Levison1994} was used.
The bodies-planetesimals that collided with the Sun or reached 2000 AU were excluded from integration.

Two models were considered. In  model $C$, bodies-planetesimals that collided with planets were excluded from integration. This model was used for studies of the number of collisions of bodies-planetesimals initially located in the zone of the terrestrial planets. 
For studies of migration to the Earth of the bodies-planetesimals initially located in the zone of the giant planets and in the outer asteroid belt, in model $MP$ at integration, planets were considered as material points, and the probabilities of collisions of bodies with planets were calculated based on the arrays of the orbital elements of migrated bodies during the considered time interval 
similar to \cite[(Ipatov \& Mather, 2004;]{IpatovMather04} \cite[Ipatov \& Mather, 2006;] {IpatovMather06} \cite[Ipatov, 2019)]{Ipatov19}. Such probabilities are relatively small. 
This model $MP$ allows to calculate probabilities of collisions of bodies-planetesimals with planets in the case of small probabilities without considering a large number of bodies.  
The probability was averaged over all considered bodies-planetesimals, and the probability per one body was considered. 
Note that this algorithm of calculations of the probabilities differed from the \"Opik's scheme (used, for example, by \cite[Morbidelli et al. (2000)]{Morbidelli2000} and \cite[Nesvorny et al. (2017)]{Nesvorny2017}). 
It included calculations of a synodic period and the region where the distance between the `first' orbit and the projection of the `second' orbit onto the plane of the `first' orbit is less than the sphere of action (i.e., the Tisserand sphere).

In each variant of calculations, 250 initial bodies-planetesimals were considered. The semi-major axes $a_o$ of the initial orbits of the bodies varied from $a_{\min}$ to $a_{\min}+d_a$. The number of bodies with $a_o$ was proportional to $a_o^{0.5}$. 
For ($i+1$)-th body-planetesimal, the value of $a_o$ was calculated with the use of the formula $a_o(i+1) =(a_{oi}^2 +[ (a_{\min} +d_a)^2 - a_{\min}^2]/N_o)^{1/2}$, 
where $a_{oi}$ is the value of $a_o$ for $i$-th body-planetesimal, and $N_o = 250$.
 Initial eccentricities  and  inclinations
of orbits of bodies equaled to $e_o$ and $i_o$=$e_o$/2 rad, respectively. 

Three main series of calculations were considered. 
In series $Ter$, bodies-planetesimals in the zone of the terrestrial planets were considered. In this case $d_a=0.2$ AU, exclusive for $a_{\min}$=1.5 AU (when $d_a=0.5$ AU). For different variants of $Ter$ calculations, the values of $a_{\min}$ varied with a step of 0.2 AU from 0.3 to 1.5 AU.  $e_o$ was equaled to 0.05 or 0.3.
In series $AsB$, initial bodies were located in the outer asteroid belt, $d_a=0.1$ AU, $a_{\min}$ varied from 3 to 4.9 AU, and $e_o$ equaled to 0.02 or 0.15. In series $JN$, initial bodies were located in the zone of the giant planets, $d_a=2.5$ AU, $a_{\min}$ varied from 2.5 to 40 AU, and $e_o$ equaled to 0.05 or 0.3.
In series $Ter$, the gravitational influence of the Sun and all planets was taken into account. In series $AsB$ and $JN$, the gravitational influence of Mercury was not considered.

\section{Mixing of planetesimals in the feeding zone of the terrestrial planets}

In \cite[(Ipatov, 2019)]{Ipatov19} the probability of a collision of a planetesimal with a planet was calculated for the  $MP$ model based on the arrays of orbital elements of bodies-planetesimals during the considered time interval. For a few planetesimals from the feeding zone of the terrestrial planets such calculated probabilities exceeded 1. Based on the arrays of orbital elements used in \cite[(Ipatov, 2019)]{Ipatov19}, 
I recalculated the probabilities of collisions of planetesimals with planets for the model in which calculations of the probability $p_i$ of a collision of each $i$-th planetesimal with a planet stopped when the probability reached 1. Though the recalculated probabilities could be smaller than those in (\cite[Ipatov, 2019]{Ipatov19}),  the conclusions about the mixing of planetesimals were similar. 
The probabilities of collisions of planetesimals with planets could be even smaller for $C$ calculations that exclude collided planetesimals from integration.

Below I present the probabilities of collisions of planetesimals with planets obtained at the $C$ calculations that exclude collided planetesimals from integration of their motion and consider present planets.
For $e_o=0.05$ during the dynamical lifetimes $T_{end}$ of planetesimals, the fraction $p_E$ of initial planetesimals collided with the Earth was about 0.35 at $a_{\min}$ equal 0.9 AU, 0.2 at $a_{\min}$ equal 0.7 or 1.1 AU, and 0.1-0.15 at $a_{\min}$ equal to 0.5, 1.3, or 1.5 AU. 
For $e_o=0.3$, the values of $p_E$ were about 0.15-0.2 for $0.7 \le a_{\min} \le 1.3$ AU, and about 0.1 at $a_{\min}$ equal to 0.5 or 1.5 AU. 
The ratio of the number of planetesimals collided with the Earth during the first $T=1$ Myr to that collided during $T_{end}$ exceed 0.5 at $a_{\min}=0.9$ AU and $e_o=0.05$. 
The similar ratio at $T=10$ Myr was between 0.4 and 0.7 at $a_{\min}$ equal to 0.5, 1.1 and 1.3 AU for $e_o$ equal to 0.05 or 0.3. The values of $T_{end}$ could exceed 1000 or 2000 Myr for $e_o=0.05$ and $ 0.5 \le a_{\min} \le 1.1$ AU. For other considered variants, they typically equaled to several hundreds of million years. 
The fraction $p_V$ of planetesimals collided with Venus exceed $p_E$ usually at all considered variants exclusive for $a_{\min}=0.9$ AU and $e_o=0.05$ and for $a_{\min}=1.5$ AU. 
For the latter variants, the ratio $p_V/p_E$ was about 0.7-0.9 at $T_{end}$.
The ratio $p_V/p_E$ was about 4-6, 2-3, 1.3-1.4, and 1.6-1.8 for $a_{\min}$ equal to 0.5, 0.7, 1.1, and 1.3 AU, respectively (both at $e_o$ equal to 0.05 or 0.3). 
The above results testify that more than a half of planetesimals with initial semi-major axes between 0.9 and 1.1 AU and initial eccentricities $e_o=0.05$ that collided with the Earth of its present mass collided in less than 1 Myr.
 The Earth embryo first mainly accumulated planetesimals that moved not far from its orbits. Planetesimals that initially had been located near orbits of other planets could collide with the Earth after tens or even hundreds of million years after the beginning of the evolution.
At the late stages of accumulation of Earth and Venus, both planets accumulated planetesimals originated at the same distance from the Sun.

\section{Migration of bodies from the zone of the giant planets to the terrestrial planets}

Sources of the water delivered to the Earth  included bodies-planetesimals migrated from the region of the outer asteroid belt and from the region beyond the Jupiter's orbit.
Migration of bodies for different values of $a_{\min}$ from 2.5 to 40 AU was studied for  $JN$ series \cite[(Ipatov, 2020)]{Ipatov2020}.
In the Solar System at $a_{\min} \le 10$ AU,  the values of the probability $p_E$ of a collision of a planetesimal with the Earth averaged over 250 bodies can vary by up to a factor of a thousand for different calculation variants with 250 planetesimals, and in some calculations they exceeded $10^{-3}$. This large difference can be caused by that in some calculations there was a planetesimal, for which the probability of a collision with the Earth was greater than for all other 249 planetesimals because one of thousands of planetesimals could get the Earth-crossing orbit and move in it for millions years. 
For comparison, the mean time for motion of a former Jupiter-crossing object in Earth-crossing orbit is about 30 Kyr.

While averaging over thousands of planetesimals, at $5 \le a_o \le 12$ AU, $p_E$ could exceed $2\cdot10^{-6}$ 
by at least a factor of several. On average, for a planetesimal at $20 \le a_o \le 40$ AU, $p_E$ was about $10^{-6}$. 
At $p_E = 2 \cdot 10^{-6}$ and the total mass of planetesimals of about $100m_E$ (where  $m_E$ is the mass of the Earth), the total mass of planetesimals collided with 
the Earth equaled to about the mass of the Earth's oceans ($\sim 2\cdot10^{-4}m_E$). 
Some fraction of material delivered from the zone of the giant planets to the Earth was composed of water and volatiles. 
In comets such fraction does not exceed 1/3 (e.g., \cite[Davidsson et al., 2016]{Davidsson2016}; \cite[Fulle et al., 2017)]{Fulle2017}. 
However, some authors suppose that primary planetesimals could contain more ice than it is now found in comets.
\cite[Lodders (2003),]{Lodders2003} \cite[Howard et al. (2014),]{Howard2014} and \cite[Ciesla et al. (2015)]{Ciesla2015}
 supposed that solids beyond the water line should be ~50 per cent water in mass.
The total mass of water delivered to the Earth from beyond Jupiter's orbit could exceed the mass of the Earth's oceans 
if the total mass of planetesimals was about $200m_E$. 
The ratio of the total mass of the material delivered from beyond the orbit of Jupiter 
to a planet to the mass of the planet was about two times greater for Mars than that for the Earth, and such ratios for 
Mercury and Venus were a little greater than that for the Earth. Some bodies from the zone of Uranus and Neptune may have fallen onto the Earth within more than twenty million years.

\section{Migration of bodies from the zone of the outer asteroid belt to the terrestrial planets}

For the series $AsB$ (with $d_a=0.1$ AU), $a_{\min}$  varied from 3 to 4.9 AU; $e_o=0.02$ and $e_o=0.15$ were taken.
At time interval $T=100$ Myr, the values of the probability $p_E$ of a collision of a planetesimal with the Earth averaged over 250 planetesimals vary from less than $10^{-6}$ 
to values of the order of $10^{-3}$ (and of 0.01 at $T=1000$ Myr) at $a_{\min}<4$ AU \cite[(Ipatov, 2021)] {Ipatov2021}.  
Generally, the values of $p_E$ are often between $10^{-6}$ and $10^{-5}$ 
at $a_{\min} \ge 4.1$ AU, as for calculations with $5 \le a_{\min} \le 10$ AU.
Calculations with 250 planetesimals with $p_E > 2\cdot 10^{-5}$ were obtained more often at $e_o=0.02$ for $3.2 \le a_{\min} \le 3.3$ AU and $a_{\min}=3.5$ AU  
(at $e_o=0.15$ also for $3.8 \le a_{\min} \le 3.9$ AU). 
In some variants of calculations at $a_{\min}=3.3$ AU and $a_{\min}=3.5$ AU for $e_o=0.02$, and at $a_{\min}$ equal to 3.0, 3.2, 3.3 and 3.8 AU  for $e_o=0.15$,
the value of $p_E$ was greater than $10^{-3}$ for a time interval up to a few billion years. 
Most of collisions with the Earth of bodies at $4 \le a_o \le 5$ AU occurred during the first 10 Myrs. 
Bodies-planetesimals that initially crossed the Jupiter's orbit could reach the Earth's orbit mostly within the first Myr 
after the formation of Jupiter.
For $3 \le a_{\min} \le 3.5$ AU and $e_o \le 0.15$, some bodies could fall onto the Earth and the Moon in a few billions years. 
For example, $p_E=4 \cdot 10^{-5}$ for $a_{\min}=3.3$ AU, $e_o=0.02$ at $500 \le t \le 800$ Myr,  
and $p_E=6\cdot10^{-6}$ at $2000 \le t \le 2500$ Myr. The zone of the outer asteroid belt can be one of the sources of the late heavy bombardment.

\section{Probabilities of collisions of migrated bodies with the Moon}

During accumulation of the terrestrial planets, the ratio of the number of planetesimals from the feeding zone of these planets colliding 
with the Earth to that colliding with the Moon varied mainly from 20 to 40. 
For bodies arriving from distances from the Sun greater than 3 AU, this ratio was close to 17. 
The ratio of the total mass of planetesimals collided with a celestial object to the mass of the object was greater for the Moon than for the Earth.
However, the fraction of the material of collided planetesimals that was left in a celestial object was greater for the Earth than that for the Moon.

The characteristic velocities of collisions of planetesimals from the feeding zone of the terrestrial planets with the Moon varied mostly from 8 to 16 km/s, 
and velocities of collisions with the Earth were from 13 to 19 km/s,
depending on the initial distances of planetesimals from the Sun and on their initial eccentricities \cite[(Marov \& Ipatov, 2021)]{Marov2021}. 
The velocities of collisions with the Moon of bodies 
that came from the feeding zones of Jupiter and Saturn were mainly from 20 to 23 km/s. For the Earth this diapason was from 23 to 26 km/s.
The characteristic velocities of collisions of the planetesimals, originally located at a distance from the Sun from 0.7 to 1.1 AU, with the embryos of the Earth and the Moon with masses 10 times less than 
the present masses of these celestial objects, were mainly in the range from 7 to 8 km/s for the embryo of the Earth and from 5 to 6 km/s for the embryo of the Moon. 

\section{Time variations in the number of 1-km near-Earth objects}

 Based on analysis of lunar craters, \cite[Mazrouei  et al. (2019)] {Mazrouei2019} concluded 
 that the probability of collisions of near-Earth objects (NEOs) with the Moon increased 2.6 times 290 Myr ago. For the model, in which the probability of a collision of a NEO with the Moon was equal to the current value for the last 290 Myr, and before that within 810 Myr it was 2.6 times less than the current value, the number of formed craters would be 0.6 from (i.e., it would be 1.7 times less than) the estimate obtained on the basis of the current number of NEOs. \cite[Ipatov et al. (2020)] {Ipatov2020}
 compared the number of lunar craters larger than 15 km across and younger than 1.1 Ga with the estimates of the number of craters that could have been formed for 1.1 Ga if the number of near-Earth objects and their orbital elements during that time were close to the corresponding current values. These estimates do not contradict to the growth in the number of near-Earth objects after probable catastrophic fragmentations of large main-belt asteroids, which may have occurred over the recent 300 Myr; however, they do not prove this increase. For some models, the cratering rate may have been constant over the recent 1.1 Ga.
The estimates made in  \cite[(Ipatov et al. 2020)] {Ipatov2020}
 allow an increase in the probability of collisions of NEOs with the Moon by a factor of 2.6 about 290 Myr ago. 
With this conclusion, the paper by \cite[Mazrouei et al (2019)] {Mazrouei2019} agrees better with the estimates based on the number of craters per unit area for the region of the Ocean of Storm and other seas of the visible side of the Moon. It was assumed that the number of Copernican craters per unit area for the entire surface of the Moon could be approximately the same as that for the region of the seas, that is, be more than the current estimate for the entire surface of the Moon. 

\section{Conclusions}
The outer layers of the Earth and Venus could accumulate similar planetesimals from different regions of the feeding zone of the terrestrial planets. 
The total mass of water delivered to the Earth from beyond Jupiter's orbit could exceed the mass of the Earth's oceans.
 The zone of the outer asteroid belt could be one of the sources of the late heavy bombardment. 
The bodies that came from the  zone of Jupiter and Saturn 
typically collided with the Earth and the Moon with velocities from 23 to 26 km/s and  from 20 to 23 km/s, respectively.

\section{Acknowledgements}
Studies of the migration of planetesimals to the terrestrial planets were carried out as a part of the state assignments of the Vernadsky Institute of RAS.
The studies of the migration of planetesimals to the Moon were supported by the grant of Russian Science Foundation N 21-17-00120.

\end{document}